\documentclass[journal]{IEEEtran}
\ifCLASSINFOpdf
\else
\fi

\usepackage[pdftex]{graphicx}
\usepackage{amsmath}
\usepackage{amssymb}
\usepackage{cite}

\begin{document}


\title{\LARGE Automatic Measurement Technique of Electromagnetic Rotation in a Nonreciprocal Medium}

\author{Swadesh~Poddar, Alexander~M.~Holmes,~\IEEEmembership{Student Member,~IEEE},
        and~George~W.~Hanson,~\IEEEmembership{Fellow,~IEEE}

\thanks{Manuscript received xxx; revised xxx.}
\thanks{S. Poddar,  (email: spoddar@uwm.edu}}

\maketitle

\begin{abstract}
This article aims at developing a simple, automated test bench procedure to measure polarization
rotation of an electromagnetic wave upon
propagation through a nonreciprocal medium. A comprehensive measurement approach is developed from the scattering matrix. The proposed
measurement procedure is demonstrated using an electronically tunable nonreciprocal
metamaterial, and the accuracy of the proposed method is  compared with the well-accepted conventional measurement technique. 
\end{abstract}

\begin{IEEEkeywords}
Faraday rotation, non-reciprocity, polarization rotation.
\end{IEEEkeywords}


\IEEEpeerreviewmaketitle

\section{Introduction}


\IEEEPARstart{P}{}olarization rotation, the change in orientation of the plane of polarization about the microwave/optical axis of a linearly polarized wave as it propagates through certain materials, is an important concept in both reciprocal and nonreciprocal systems. The precise determination  of the polarization state of an electromagnetic wave is fundamental for a wide variety of applications in microwaves, radio frequency front ends, astronomy, optics and terahertz technology, and remote sensing, to name just a few. A linearly polarized electromagnetic wave may be represented as the sum of two circularly polarized waves (right-and left-handed circular polarized waves) with opposite handedness and equal amplitude. 
Faraday rotation occurs when electromagnetic waves propagate through a nonreciprocal medium, whereupon the right-handed and left-handed circularly polarized waves (L/RHCP) propagate with different phase velocities, and experience different impedance and refractive indices.
This LHCP and RHCP waves start to accumulate a phase difference as they propagate through the medium, therefore, with respect to time and distance, the direction of linear polarization
changes. The greater the deviation in phase velocity, the greater the rotation \cite{lax1962microwave, asadchy2020tutorial, caloz2018nonreciprocity}.

The robust practical  measurement of polarization rotation in a nonreciprocal medium is very challenging. Various approaches have been investigated to accurately measure Faraday rotation in various frequency regimes \cite{ FR6, FR1, caloz2005electromagnetic, FR3, FR2, FR4} and the most conventional approach is to measure polarization rotation by rotating an antenna \cite{Kodera4}. However, manual approaches to measure Faraday rotation lack precision and are time consuming. Therefore, a simple, realizable, cost effective, accurate test-bench model is of immense importance to measure Faraday rotation. In this work, we have developed a simplified and automatic S-parameter translation to measure polarization rotation in a nonreciprocal environment using a network analyzer. We apply our proposed measurement method to one of our previously developed nonreciprocal metamaterials and we observe good agreement with manual measurement. Furthermore, our proposed model will work for a non reciprocal slab of finite thickness.  

The paper is organized as follows. In Section II, we provide insight to the mechanism of EM nonreciprocity and various polarization rotations, in Section III, we present the S parameter analysis leading to simplified automatic Faraday rotation measurement and, in Section IV, we benchmark our proposed measurement model with a conventional approach on one of our previously implemented tunable nonreciprocal meta-surfaces.



\section{Electromagnetic Non-reciprocity and Polarization Rotation}

A system can be subdivided into two broad categories based on its interaction with electromagnetic waves: reciprocal and non-reciprocal. A reciprocal system is one in which the fields created by the source at the observation point is the same when source and observation points exchange position, which can be represented by the scattering parameter equality $\mathbf{S} = \mathbf{S^T}$. Devices such as antennas, passive RLC electrical circuits, and most RF components are reciprocal. On the other hand, nonreciprocity, achieved by breaking time reversal symmetry, dictates that the field created
by the source and measured at the observation point is different when source and observation points are interchanged, and can be represented by $\mathbf{S}$ $\neq$ $\mathbf{S^T}$ \cite{Nagulu2020, kord, caloz2018nonreciprocity}. As an example, nonreciprocal components such as gyrators, isolators, and circulators, are very important because of their impact in military applications, telecommunications and so on \cite{Nagulu2020, ULMKodera}. There are various reported approaches to achieve nonreciprocity since the 1950s such as using ferrite materials biased by a static magnetic field, and the latest concept of achieving nonreciprocity using time modulation \cite{kord}. In all reported cases of achieving nonreciprocity, the fundamental concept is breaking time reversal symmetry \cite{caloz2018nonreciprocity, asadchy2020tutorial}.

\begin{figure}[!t]
\centering
\includegraphics[width=1\columnwidth]{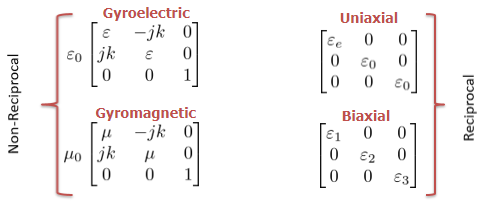}
\caption{Matrix representation of various media and classification of non-reciprocal and reciprocal media }
\label{fig:Equations}
\end{figure}

Polarization rotation is directly related to material properties, and system conditions. For example, polarization rotation in various material and structural systems, such as chiral materials, wire grid polarizers, ferrites, plasmas, birefringents, and gyro-tropic mediums are different. If a linearly polarized wave travels forward through a nonreciprocal medium by a distance $l$, gets reflected, and travels
back to the starting point, the polarization rotation angle can be represented as \cite{orfanidis2016electromagnetic}. 

\begin{equation}
\phi = \frac{1}{2}(k_+ - k_-)l,
\label{eq:gyromagnetic_1}
\end{equation}
where wavenumber is $k_\pm = \omega\sqrt{\epsilon\mu_{\pm}}$, impedance is $\eta_{\pm} = \sqrt{\frac{\mu_{\pm}}{\epsilon}}$ and permeability is $\mu_{\pm} = \mu_1\pm\mu_2$. Therefore, the LHCP and RHCP waves exhibit different phases, impedances, and refractive indices.

\begin{figure}[!b]
\centering
\includegraphics[width=1.2\columnwidth]{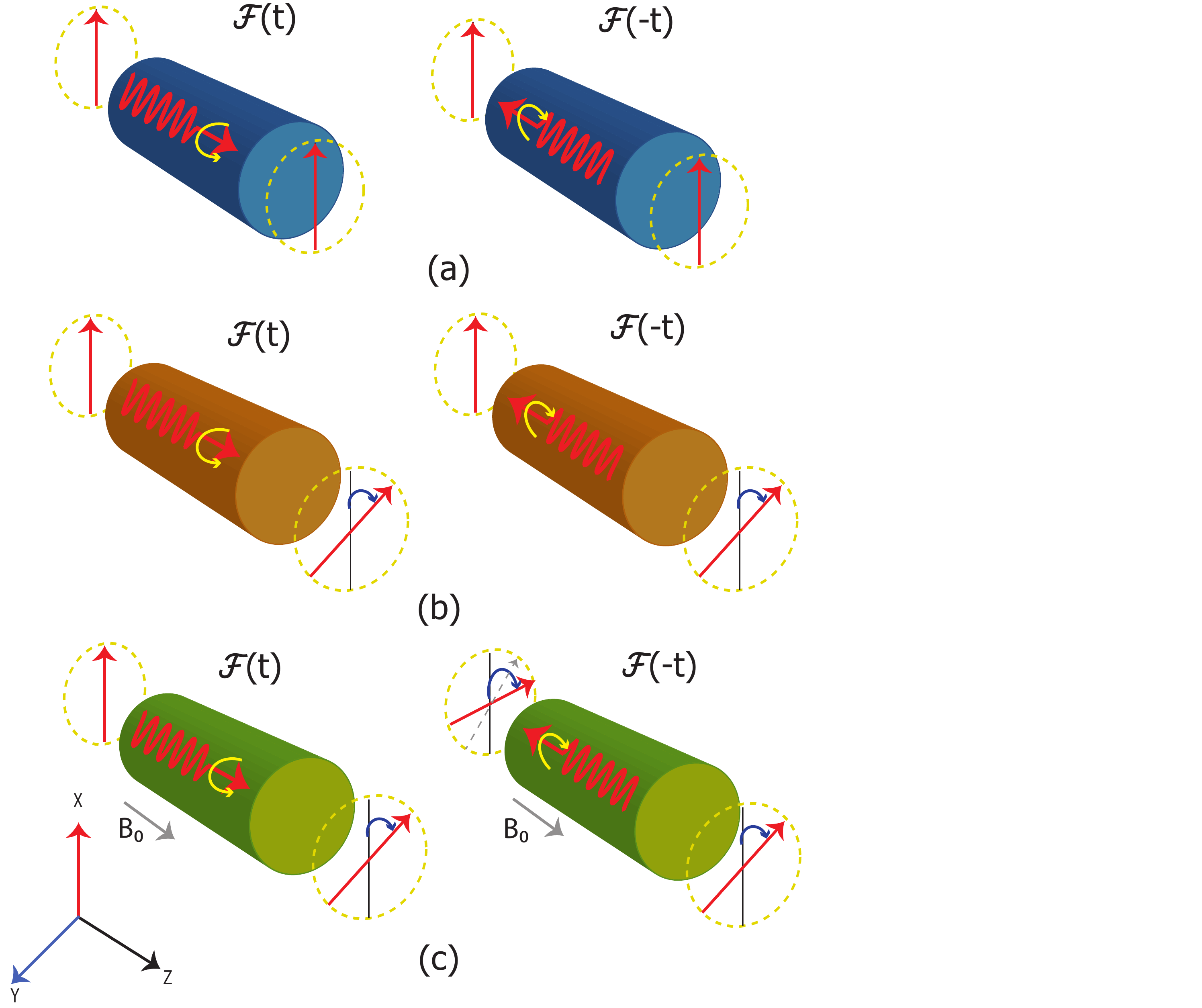}
\caption{Criteria for time reversal symmetry and assymmetry. (a) Isotropic medium (b) Chiral medium (Reciprocal). (c) Faraday system by unaltered external magnetic bias, hence TRS breaking, non-reciprocal system.}
\label{fig:TRS}
\end{figure}

Polarization rotation, the phase difference between cross polarized components, and transmission parameters ($S_{21}$, $S_{12}$) are key performance indicators to characterize nonreciprocity. For example, nonreciprocal gyrotropy is the usual way to achieve nonreciprocity via having a certain type of off-diagonal elements in the $\mu$ or $\epsilon$ tensor. Figure \ref{fig:Equations} represents the matrix form of various reciprocal and nonreciprocal cases. In bulk media nonreciprocity requires asymmetric electric permittivity and magnetic permeability which can be classified as gyroelectric (${\varepsilon} \neq \mathbf{{\varepsilon}^{T}}$)  and gyromagnetic (${\mu} \neq \mathbf{{\mu}^{T}}$), respectively where in Fig.\ref{fig:Equations}, k in the off-diagonal terms represent the gyrotropic parameters. Nonreciprocal gyrotropic and gyromagnetic materials exhibit off diagonal tensor elements, whereas, reciprocal cases such as uniaxial or biaxial material don't have any off diagonal elements, hence gyrotropic response of the magnetized ferrites is evident. In addition because of off diagonal tensor components, nonreciprocal materials exhibit cross polarized phase and magnitude difference \cite{Swadesh}. It is important to note here that systems with symmetric tensors ${\varepsilon} = \mathbf{{\varepsilon}^{T}}$ or ${\mu} = \mathbf{{\mu}^{T}}$ with nonzero off-diagonal elements exhibit polarization rotation (optical activity), however, in a  reciprocal way \cite{krasnok}. 

To characterize material interactions with electromagnetic waves, in Fig. \ref{fig:TRS} we show three different systems that can be classified with respect to time reversal symmetry and asymmetry \cite{caloz2018nonreciprocity}. Figure \ref{fig:TRS} (a) defines an isotropic medium where there is not any polarization rotation and hence, is classified as reciprocal. Figure \ref{fig:TRS} (b) represents a chiral system, where the x polarized wave travels through the medium and experiences a polarization rotation of $\phi^0$ at the receiver end, and in the reverse direction it will experience a polarization rotation of $-\phi^0$. As a result, the field polarization of a chiral system symmetrically returns
to its original state, hence, this is a reciprocal system. 
Figure \ref{fig:TRS} (c) depicts Faraday rotation in the presence of a static $\textbf{B}_{0}$ field. The applied static magnetic bias breaks time-reversal symmetry, making the process time-reversal asymmetric \cite{caloz2018nonreciprocity, caloz2018nonreciprocityII, asadchy2020tutorial}. In comparison to the chiral case, the wave experiences $\phi^0$ rotation, then in the reverse direction it experiences another $\phi^0$ rotation, revealing Faraday rotation \cite{ caloz2019electromagnetic}. To recapitulate, the process of polarization rotation can be reciprocal or nonreciprocal. In the case of Faraday rotation the sign (clockwise or anticlockwise) is determined relative to the axis along which time-reversal (TR) symmetry is broken, whereas in chiral media exhibits reciprocal magneto-electric coupling, the sign of rotation is determined relative to the propagation direction \cite{Floess2015, Swadesh}. The next section will be focused on developing a realistic model for the measurement of polarization rotation in a transparent medium.

\begin{figure}[htbp]
\centering
\includegraphics[width=1\columnwidth]{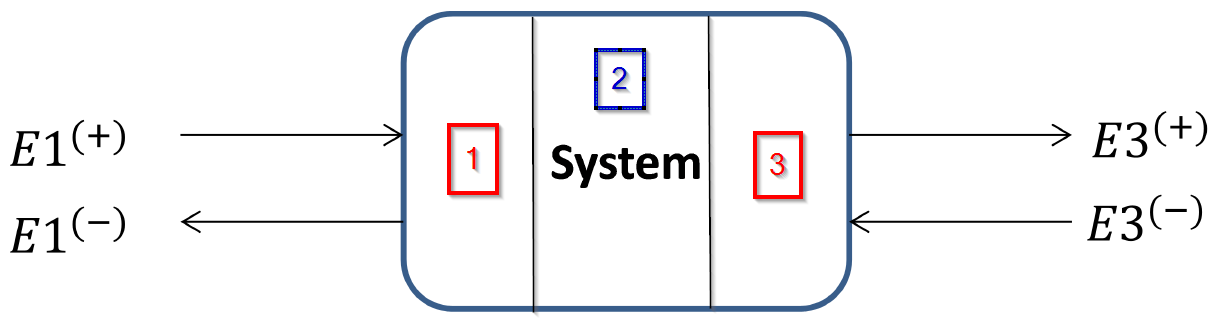}
\caption{A two port network model where medium 1 and 3 are reciprocal and medium 2 is nonreciprocal.}
\label{fig:system_setup}
\end{figure}

\section{Principle of Operation and Practical Realization}

\begin{figure*}[!t]
\centering
\includegraphics[width=2\columnwidth]{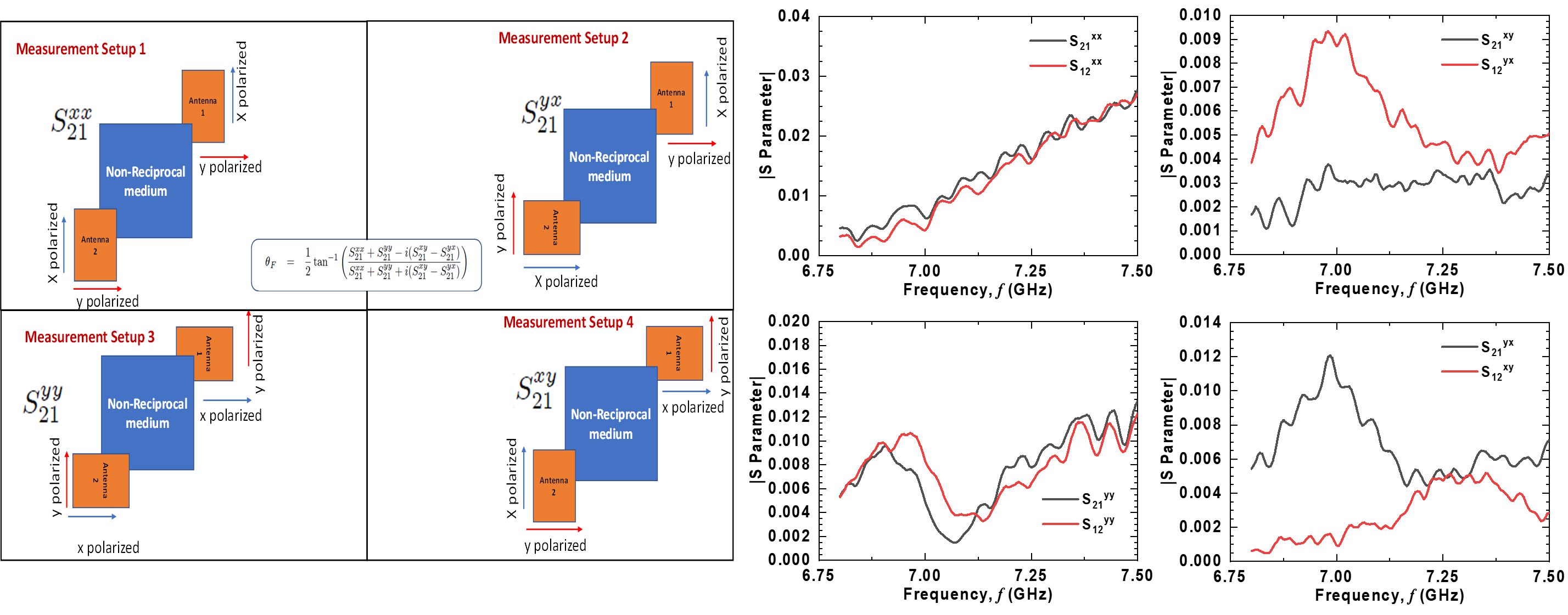}
\caption{(left panel) Automatic Faraday rotation measurement in 4 steps leading to combined numerical calculation. (right panel a,b,c,d) Measured S-parameter spectra for Co-Pol and Cross-Pol configurations used to calculate Faraday rotation.}
\label{Meas_S_Para_Auto}
\end{figure*}


Figure \ref{fig:system_setup} depicts a two-port network segmented into three parts used for our mathematical modeling of Faraday rotation, where segments 1 and 3 represent reciprocal media and segment 3 represents a nonreciprocal medium. Superscript +/- represents forward/backward waves. For practical realization, focus will be on measuring $E_1^{+}$, $E_1^{-}$, $E_3^{+}$ and $E_3^{-}$ in terms of S matrices which can be represented by reflection and transmission components as%
\begin{equation}
\overrightarrow{\mathbf{S}} = \begin{bmatrix}
\overrightarrow{\mathbf{r}}^{(-)} & \overrightarrow{\mathbf{t}}^{(-)} \\ 
\overrightarrow{\mathbf{t}}^{(+)} & \overrightarrow{\mathbf{r}}^{(+)}
\end{bmatrix} = \begin{bmatrix}
\overrightarrow{\mathbf{S}}_{11} & \overrightarrow{\mathbf{S}}_{12} \\ 
\overrightarrow{\mathbf{S}}_{21} & \overrightarrow{\mathbf{S}}_{22}
\end{bmatrix},
\label{Eq:S_matrix}
\end{equation}%
where in terms of medium 1, 2 and 3, each component can be described as%
\begin{equation}
\begin{split}
   \overrightarrow{\mathbf{S}}_{11}=\overrightarrow{\mathbf{r}}(1\rightarrow{2}\rightarrow{1})= \overrightarrow{\mathbf{r}}^{(-)}\\
      \overrightarrow{\mathbf{S}}_{12}=\overrightarrow{\mathbf{t}}(1\rightarrow{2}\rightarrow{3})= \overrightarrow{\mathbf{t}}^{(+)}\\
            \overrightarrow{\mathbf{S}}_{21}=\overrightarrow{\mathbf{t}}(3\rightarrow{2}\rightarrow{1})= \overrightarrow{\mathbf{t}}^{(-)}\\
                        \overrightarrow{\mathbf{S}}_{22}=\overrightarrow{\mathbf{r}}(3\rightarrow{2}\rightarrow{3})= \overrightarrow{\mathbf{r}}^{(+)} 
   \end{split}
\end{equation}%
Therefore, Eq. \ref{Eq:S_matrix} can be represented as 
\begin{equation}
\overrightarrow{\mathbf{S}}=
\begin{bmatrix}
 S^{yy}_{11}&  S^{xy}_{11} & S^{yy}_{12}  &  S^{xy}_{12}\\ 
 S^{yx}_{11}&  S^{xx}_{11} & S^{yx}_{12}  &  S^{yy}_{12}\\ 
 S^{yy}_{21}&  S^{xy}_{21} & S^{yy}_{RR}  &  S^{xy}_{22}\\ 
 S^{yx}_{21}&  S^{xx}_{21} & S^{yx}_{22}  &  S^{yy}_{22}
\end{bmatrix},
\label{eq:S_para}
\end{equation}%
where the wave is being transmitted in the Z direction and the first and second subscript/superscript represents port number/polarization of receiver and transmitter antenna, respectively. We obtain Faraday rotation when the cross polarized transmission component from port 1 and port 2 are not the same. We can easily calculate Faraday rotation by using Eq. \ref{eq:Faraday} as described in the appendix \cite{lax1962microwave, shivola} %
\begin{equation}
\theta _{F}=\frac{1}{2}\tan ^{-1}(\frac{t_{\circlearrowleft \circlearrowleft
}}{t_{\circlearrowright \circlearrowright }}),
\label{eq:Faraday}
\end{equation} 
where $t_{\circlearrowleft \circlearrowleft
}$ defines the transmission component due to both incident and received waves being LHCP. Similarly, $t_{\circlearrowright \circlearrowright
}$ defines the transmission component due to an incident and received waves being RHCP. Therefore, from Eq. \ref{eq:S_para} Faraday rotation can be represented in terms of S parameters in linear polarization as
\begin{equation}
\theta _{F}=\frac{1}{2}\tan ^{-1}\left ( \frac{S_{21}^{xx}+S_{21}^{yy}-i(S_{21}^{xy}-S_{21}^{yx})}{S_{21}^{xx}+S_{21}^{yy}+i(S_{21}^{xy}-S_{21}^{yx})} \right ).
\label{eq:Faraday_S}
\end{equation}

\section{Simulation and Measurement}

In our previous work \cite{meta}, we calculated and measured the Faraday rotation of a normally incident plane wave on an electronically tunable non-reciprocal metasurface. Tunable non-reciprocity in the metasurface was introduced using a field effect transistor (FET) with variable gate-source and drain-source bias conditions. Here, we elaborate on the measurement procedure used in \cite{meta}, and show agreement between the calculated and manually measured results for a set bias condition where the drain-source current is saturated.

\begin{figure}[!b]
\centering
\includegraphics[height=0.2\textheight]{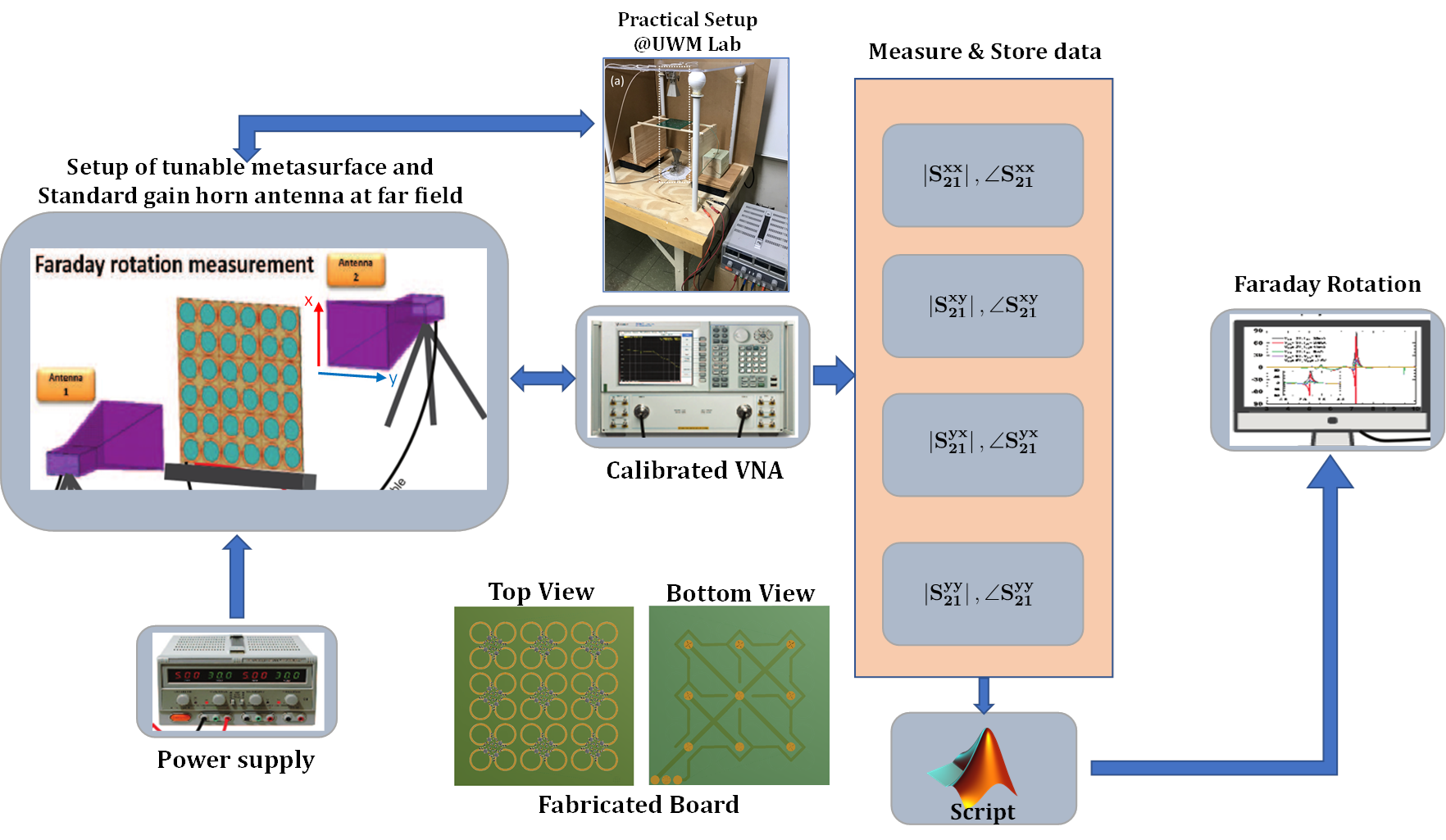}
\caption{Process flow of proposed automatic Faraday rotation as a function of frequency}
\label{fig:Process-flow}
\end{figure}

Figure \ref{fig:Process-flow} shows a block diagram of the measurement procedure to calculate Faraday rotation from the measured S-parameters over the operational frequency range. The diagram depicts the fabricated metasurface placed equidistant between the antennas, in the far field. A 30V 5A HY3005F-3 linear DC power supply is used to bias the metasurface. Once the the mechanical setup and calibration (S-Short, O-Open, L-Load, T-Through) of the Keysight E8361 vector network analyzer (VNA) is completed, a measurement is performed to obtain the co-polarized and cross-polarized S-parameter coefficients, which are then fed into Eq. (\ref{eq:Faraday_S}) to calculate Faraday rotation with respect to operating frequency. In our proposed transmission based measurement, four S-parameter coefficients (i.e., $S_{21}$) are measured for different transmit and revieve polarizations. For example, the cross-polarized S-parameter $S_{21}^{xy}$ ($S_{21}^{yx}$) is described as the transmission coefficient from antenna 1 to antenna 2 when the incident wave is y (x) polarized and the received wave is x (y) polarized. Similarly, the co-polarized S-parameter $S_{21}^{yy}$ ($S_{21}^{xx}$) is described as the transmission coefficient from antenna 1 to antenna 2 when the incident wave is y (x) polarized and the received wave is y (x) polarized. A non-reciprocal system will exhibit a phase difference in the cross-polarization, however, the co-polarized states exhibit no phase difference. This principle holds for both Faraday (transmission based) and Kerr (reflection based) rotation 
\cite{meta,Swadesh}.

\begin{figure}[htbp]
\centering
\includegraphics[width=1\columnwidth]{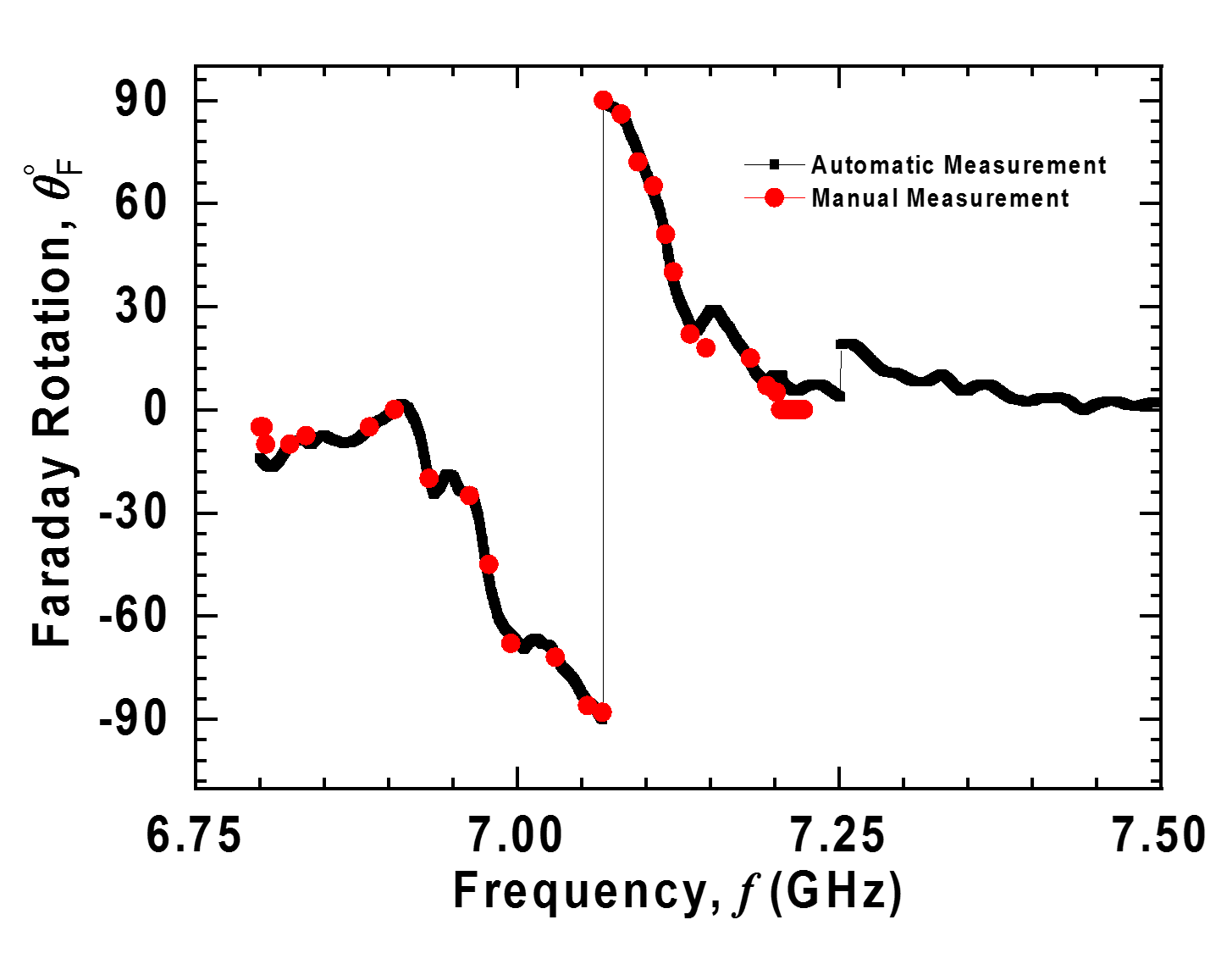}
\caption{Measurement of Faraday rotation based on automatic and manual process }
\label{fig:Measurement_FR}
\end{figure}

Figure \ref{Meas_S_Para_Auto}(a)-(d) depicts the various co-polarized and cross-polarized states of the receiver with respect to the transmitter, while Fig. \ref{Meas_S_Para_Auto}(e)-(h) shows the measured S-parameters associated with each case in (a)-(d), respectively. From the measurement, it is clear that the co-polarized measurements (a)-(b) show very little sign of isolation while the cross-polarized measurements (c)-(d) exhibit a large amount of isolation at the resonant frequency. A major challenge in measuring Faraday rotation is that while transmission is maximum at resonance, it is still very low. To account for Faraday rotation accurately, the signal from antenna 1 to antenna 2 and vice versa must pass through the metasurface only. As a result the metasurface at far field distance from both antennas must be sufficiently large. If the board is not large enough relative to the antenna beamwidth at the metasurface position, a PEC material at the operating frequency can be used surrounding the metamaterial to block any direct signal transmission from antenna 1 to antenna 2. Once the co-polarized and cross-polarized S-parameter coefficients are measured, an in house script is used to perform the Faraday rotation calculation with respect to frequency.

For comparison, a manual method to measure Faraday rotation at different operating frequencies points can be used (see Ref. \cite{meta} for details regarding the measurement setup). Figure \ref{fig:Measurement_FR} shows the measured Faraday rotation obtained using the automatic (calculated) and manual measurement with respect to the frequency range of operation. To manually measure Faraday rotation, the receive antenna is rotated manually until maximum transmission (i.e., maximum $\left| S_{21} \right|$ or $\left| S_{12} \right|$) is achieved.


\begin{table}[ht]
\caption{COMPARISON OF THE RECENTLY PROPOSED MEASUREMENT OF POLARIZATION ROTATION}
\begin{tabular}{ccccc}
\hline\hline 
Criteria & \cite{FR5} & \cite{Kodera4} & \cite{FR2} & This Work\\ [0.5ex] 
\hline
\begin{tabular}[c]{@{}c@{}}Measurement\\ Type\end{tabular}       & Discrete                                              & Discrete                                              & Continuous                                                        & Continuous                                            \\
Setup                                                            & \begin{tabular}[c]{@{}c@{}}Open \\ Space\end{tabular} & \begin{tabular}[c]{@{}c@{}}Open \\ Space\end{tabular} & Waveguide                                                         & \begin{tabular}[c]{@{}c@{}}Open \\ Space\end{tabular} \\
\begin{tabular}[c]{@{}c@{}}Measured \\ Component\end{tabular}    & Tranmission                                           & Reflection                                            & Transmission                                                      & Transmission                                          \\
\begin{tabular}[c]{@{}c@{}}Measured \\ Co-Cross Pol\end{tabular} & No                                                    & No                                                    & \begin{tabular}[c]{@{}c@{}}Not \\ Reported\end{tabular}           & Yes                                                   \\
Process                                                          & Manual                                                & Manual                                                & \begin{tabular}[c]{@{}c@{}}Automatic\\ (Backfitting)\end{tabular} & Automatic                                             \\
Tunability                                                       & Partial                                               & Partial                                               & Full Scale                                                        & Full Scale                                            \\
Bias                                                             & Electronic                                            & Electronic                                            & Magnetic                                                          & Electronic                                           
\end{tabular}
\label{Table 1}
\end{table}

Table 1 shows the comparison between the proposed automated system and the manual rotation measurement. In comparison with other works, our proposed measurement procedure offers a very simple, cost effective, precise measurement of polarization rotation. In addition, in our proposed model we can calculate Faraday rotation over all frequencies whereas most of the conventional approaches are based on manual measurement of Faraday rotation by rotating the antenna at various frequency points and post interpolation between those discrete data points.

\section{Conclusion}
To summarize, we have presented a S parameter model leading to a very simple and accurate test-bench setup to measure polarization rotation in an anisotropic media using a nonreciprocal metasurface. The numerical modelling has been analyzed and verified. We envision that this proposed model will add a new building block in the measurement of polarization rotation and can be implemented in any types of polarization rotation system utilizing two linearly-polarized antennas and a network analyzer.

\appendix

\section{Mapping between Transmission Matrix in linear and circular basis}
The transmission matrix in a linear basis can be mapped to a transmission matrix in a circular basis by following the advanced Jones calculus \cite{PhysRevA.82.053811} as%
\begin{equation}
\left[ 
\begin{array}{cc}
\mathbf{t}_{\circlearrowright \circlearrowright } & \mathbf{t}%
_{\circlearrowright \circlearrowleft } \\ 
\mathbf{t}_{\circlearrowleft \circlearrowright } & \mathbf{t}%
_{\circlearrowleft \circlearrowleft }%
\end{array}%
\right] 
=\frac{1}{2}\left[ 
\begin{array}{cc}
P & Q \\ 
R & S%
\end{array}%
\right],
\label{Eq:T_matrix}
\end{equation}
where\\ 
$P =t_{xx}+t_{yy}+i(t_{xy}-t_{yx})$\\ $Q=t_{xx}-t_{yy}-i(t_{xy}+t_{yx})$\\ $R=t_{xx}-t_{yy}+i(t_{xy}\text{ }+\text{ }t_{yx})$\\ $S=t_{xx}+t_{yy}-i(t_{xy}-t_{yx})$\\ and $\mathbf{t}_{\circlearrowright \circlearrowright}$=Incident RHCP to recieved RHCP, $\mathbf{t}_{\circlearrowright \circlearrowleft }$=Incident LHCP to recieved RHCP, $\mathbf{t}_{\circlearrowleft\circlearrowright}$=Incident RHCP to
recieved LHCP, and $\mathbf{t}_{\circlearrowleft \circlearrowleft}$=Incident LHCP to
recieved LHCP.

\section*{Acknowledgment}
We thank Dr. Md. Tanvir Hasan and Ragib Shakil Rafi for helpful discussions and guidance.
\ifCLASSOPTIONcaptionsoff
  \newpage
\fi
\bibliographystyle{IEEEtran}
\bibliography{bibliography}

\end{document}